\def\dodoh{7.26}
\def\dodm91{7.35}
\def\dodmax{7.50}

\def\dodohSTM{7.41}

\def\TeHummer{20000}
         \def\cred{0.335}           \def\ecred{0.061}
     \def\credGamma{0.67}       \def\ecredGamma{0.82}
\def\credRegression{0.36}  \def\ecredRegression{0.05}

\def\rduetre{3.12}
\def\erduetre{0.26}
\def\rduetremin{0.46}
\def\rduetremax{0.53}

\def\dodOHrMin{7.19} 
\def\dodOHrMax{7.28}

  \def\oiiHii{1.09}      \def\eoiiHii{0.17}
 \def\oiiiHii{0.73}    \def\eoiiiHii{ 0.01} 
 \def\oiioiii{1.82}     \def\eoiioiii{0.18}
\documentclass{aa} 
\usepackage{graphicx}
\newenvironment{deluxetable}[1]
{\def\ptwidth{70mm}
\begin{table}\caption[]{\ptcaption}
\begin{flushleft}\edef\tableformat{\string#1}\ptcolsep
\begin{tabular}{\tableformat}}{\noalign{\smallskip}\hline
\noalign{\medskip}\end{tabular}
\\\medskip\ptnotes\ptcomments\ptrefs\end{flushleft} 
\end{table}}

\newenvironment{deluxetable*}[1]
{\def\ptwidth{160.6mm}
\begin{table*}\caption[]{\ptcaption}
\begin{flushleft}\edef\tableformat{\string#1}\ptcolsep
\begin{tabular}{\tableformat}}{\noalign{\smallskip}\hline
\noalign{\medskip}\end{tabular}
\\\medskip\ptnotes\ptcomments\ptrefs\end{flushleft} 
\end{table*}}

 \def\ptcolsep{\relax}
\def\tablecaption#1{\gdef\ptcaption{#1}
                    \gdef\ptrefs{} \gdef\ptnotes{} \gdef\ptcomments{} 
                   \typeout{-- Warning: use Tablecaption first in table --}
                   } 
                   \def\ptcaption{\relax}
                   \def\ptrefs{\relax} \def\ptnotes{\relax} \def\ptcomments{\relax}

 \def\ptcomments{\relax}

 \def\ptrefs{\relax}

\def\tablenotetext#1{\gdef\ptnotes{
                                   \begin{minipage}[t]{\ptwidth}
                                     {#1}
                                   \end{minipage}
                                   \null\medskip
           \typeout{-- Warning: non-standard syntax for Tablenotetext  --}
                                  }
                    } \def\ptnotes{\relax}

\newcommand{\tablehead}[1]{\hline\noalign{\smallskip}#1\\}
\newcommand{\colhead}[1]{\multicolumn{1}{c}{#1}}
\newcommand{\startdata}{\noalign{\smallskip}\hline\noalign{\smallskip}}

\def\nodata{~--}

\newcommand{\tablenotemark}[1]{$^{\mathrm{#1}}$}

\newcommand{\tablewidth}[1]{}

\usepackage[dvips]{color}

\newcommand{\realfigure}[3]{
              \begin{figure}
              \centerline{\includegraphics[width=8.8cm]{#1}}
              \caption{#2}\label{#3}
              \end{figure}}

\newcommand{\uks}{UKS\,1927--177}
\newcommand{\efosc}{{\sc efosc2}}
\newcommand{\remp}{R_{ 23}}
\newcommand{\changed}{}                         

\newcommand{\aap}{A\&A} 
\newcommand{\aapr}{A\&A Rev.} 
\newcommand{\apj}{ApJ} 
\newcommand{\apjl}{ApJ} 
\newcommand{\aj}{AJ} 
\newcommand{\pasp}{PASP} 
\newcommand{\mnras}{MNRAS}

\newcommand{\hii}{\hbox{H{\sc ii}}}

\newcommand{\otwo}{\hbox{O{\sc ii}}}
\newcommand{\otre}{\hbox{O{\sc iii}}}
\newcommand{\netre}{\hbox{Ne{\sc iii}}}
\newcommand{\ntwo}{\hbox{N{\sc ii}}}

\begin{document}

\title{New abundance measurements in \uks, 
a very metal-poor galaxy in the Local Group
\thanks{Based on data collected at the European Southern Observatory,
         La Silla, Chile, prop. No. 63.N-0726B}
}

\author{
I. Saviane\inst{1}
\and 
L. Rizzi\inst{2,3}
\and 
E. V. Held\inst{2}
\and 
F. Bresolin\inst{4}
\and 
Y. Momany\inst{3}
}
\offprints {I. Saviane}

\institute{
European Southern Observatory, Casilla 19001,
Santiago 19, Chile \\
\email{isaviane@eso.org} 
\and
Osservatorio Astronomico di
Padova, vicolo dell'Osservatorio 5, I-35122 Padova, Italy \\
\email{(rizzi, held)@pd.astro.it}
\and
Dipartimento di Astronomia, Universit\`a di
Padova, vicolo dell'Osservatorio 2, I-35122 Padova, Italy \\
\email{momany@pd.astro.it}
\and
Institute for Astronomy, University of Hawaii, 
2680 Woodlawn Drive, Honolulu, HI 96822 \\
\email{bresolin@ifa.hawaii.edu}
}
\date {Accepted 17 May 2002}
\titlerunning{New abundance measurements in \uks}

\abstract{
We present new results from optical spectroscopy of the brightest
H{\sc ii} region in the dwarf irregular galaxy \uks\ in Sagittarius
(SagDIG).  From high signal-to-noise spectra, reddening-corrected line
flux ratios have been measured with typical uncertainties of a few
percent,
from which the oxygen abundance is rediscussed, and new abundance
estimates are derived for N and Ne.
The O abundance in SagDIG, estimated with the empirical abundance
indicator $\remp$ and other methods, is in the range 
\( 12+\log ({\rm O/H})=\dodoh \) to \( \dodmax \).
The fact that SagDIG is \( \sim 10 \) times closer than I\,Zw\,18 makes
it an ideal target to test the hypothesis of the existence of young
galaxies in the present-day universe. Indeed, stellar photometry
suggests that this galaxy may harbor a stellar population older than a
few Gyr, and possibly an old stellar component as well. The case of
SagDIG therefore supports the view that very low chemical abundances
can be maintained throughout the life of a dwarf stellar system, even
in the presence of multiple star formation episodes.
\keywords{galaxies: individual (SagDIG)
-- galaxies: abundances -- galaxies: dwarf -- galaxies: ISM --
galaxies: irregular -- Local Group}
}

\maketitle

\section{Introduction}

The quest for very metal-poor galaxies started soon after the Searle and
Sargent (\cite{ssargent72}) discovery that the interstellar medium (ISM)
of I\,Zw\,18, a blue compact dwarf galaxy, has an oxygen abundance that is
\( \sim 1/50 \) 
of the solar one.
%
%
As Searle \& Sargent (\cite{ssargent72}) pointed out, this is compatible
with two scenarios: either the galaxy is now experiencing its first
star-formation (SF) episode, or, if it contains \( \sim 10\, \rm Gyr \)
old stars, then its SF proceeds in strong bursts separated by quiescent
phases. The first hypothesis (that of a \emph{primordial} galaxy) is
relevant in the context of cosmological models that assume hierarchical
formation of structures in a cold dark matter  dominated
Universe. In that scenario, dwarf dark matter halos (\( \sim 10^{8}\, \rm
M_{\odot } \)) would be the
\emph{first} objects 
to condense out of the Hubble flow, so it is important to understand
whether they already contained stars before merging into larger
galaxies. As a first step, we can start by answering a simple
question: is there any galaxy with a very metal-poor young population
and, at the same time, a \( >10\, \rm Gyr \) old generation of stars?
Such a finding would lead to the conclusion that very low metal
abundances do not necessarily imply a very young age.

Although a red population was eventually revealed in I\,Zw\,18 
by Aloisi et al. (\cite{aloisi_etal99}), 
{\changed their data only allow for a look-back time}
of \( \sim 1 \)~Gyr, due to the large distance of the galaxy.
Izotov et al. (\cite{izotov_etal00}) argue that \( 100\, \rm Myr\) 
is a more realistic limit. As a consequence, active searches are
carried out to find closer and comparably metal-poor galaxies, in
order to have better chances to detect a cosmologically old
population. In particular, \emph{low luminosity} dwarf irregulars are
promising candidates. It is still debated whether a definite
luminosity-metallicity relation exists for dIrrs (Hidalgo-Gamez \&
Olofsson \cite{h-go98}; Hunter \& Hoffman \cite{hh99}), but it is
nevertheless reasonable to expect that chemical abundances will be
lower in low-luminosity galaxies.

Indeed, some new metal-poor, low-luminosity objects turned up in the
past as a result of 
similar efforts. The situation has been conveniently
summarized by Kniazev et al. (\cite{kniazev00}), who list two galaxies
(besides I\,Zw\,18) at \( 12+\log (\rm O/H)\leq 7.35 \).  A few months
later van Zee (\cite{vanZee00}) added UGCA~292 to the list of
metal-poor galaxies, quoting \( 12+\log ({\rm O/H})=7.30\pm 0.05 \),
and pointed out that Leo~A should be added to the list as well.

Stimulated by all these promising results, we turned our attention to
\uks\ (SagDIG), since (a) its low luminosity is in the range of
other metal-poor galaxies, 
(b) there was some indication that the oxygen abundance obtained by
Skillman et al. 
(\cite{stm89}, hereafter STM89), 
\( 12+\log ({\rm O/H})= \dodohSTM \), could be revised
downwards (see below), and (c) the other metal-poor galaxies (with the
exception of Leo~A) are more distant than \( 3\, \rm Mpc \). SagDIG
is more than two magnitudes closer in distance modulus (thus its HB is
within reach of HST observations in reasonable observing time).  

The galaxy was discovered in 1977 by Cesarsky et
al. (\cite{cesarsky_etal77}), and a shallow CMD was presented in Cook
(\cite{cook87}). A CCD survey of its \hii\ regions was carried out by
Strobel et al. (\cite{strobelEtal91}; SHK91), using the
$2.1~\rm m$ telescope at Kitt Peak.  Three objects were detected in a
field of view of $\sim 2\farcm 5 \times 2\farcm 5$, with 
H$\alpha$ fluxes of $28.1\times 10^{-15}$, $20.0\times 10^{-15}$, and
$4.09\times 10^{-15}~\rm erg\,cm^{-2}\,s^{-1}$.  Optical
spectrophotometry of the brightest object (object \#3 in Table~8 of
SHK91) was obtained by STM89, and an oxygen abundance
\( \sim 1/30 \)  of the 
solar value\footnote{{\changed In this paper we adopt the classical
value $12+\log(\rm O/H) = 8.87$ from Anders \& Grevesse 
(\cite{ande+grev89}). However, the solar oxygen abundance is
currently debated, and a lower value might be appropriate (Holweger
\cite{holweger01}; Allende
Prieto et al. \cite{alle+01}).  }}
could be determined, i.e. $12+\log(\rm O/H) \simeq \dodohSTM $. 
Radio maps of the galaxy exist as well. Young \& Lo (\cite{youngLo97})
presented deep VLA observations of the H{\sc i} in SagDIG. An amount
of \( 1.3\times 10^{7} \) solar masses of H+He were detected, forming
a circular ring which extends \( \sim 3 \) times farther than the 
optical extent of the star-forming regions.

The compactness of the galaxy, its low Galactic latitude,
and the high foreground contamination had discouraged further broadband
photometry until recently. In the last two years, imaging of this galaxy
was presented by 
Karachentsev et al. (\cite{kara_etal99}) and 
Lee \& Kim (\cite{lee_kim00}), who established the distance modulus at
$(m-M)_0=25.13\pm 0.25$, and the RGB stellar metallicity at extremely
low values ([Fe/H]$\la -2.45$).
From surface photometry, Karachentsev et
al. (\cite{kara_etal99}) found an integrated magnitude of
$M_V=-11.74$, and with a synthetic CMD analysis, they concluded that
the galaxy is currently experiencing a burst of star formation, the
star formation rate (SFR) being $10$ times larger now than the life
average.

A new imaging study of SagDIG is presented in a companion paper
(Momany et al. \cite{yaz_sagdig}). The most important results of our
photometry is represented by an upward revision of the stellar
metallicity of the galaxy and our analysis of the spatial distribution
of different stellar populations as a function of age.  Assuming a low
reddening for old RGB stars, we derived [Fe/H]$=-2.08\pm 0.20$ for the
red giant branch stars. 

In this paper we present our new determination of the oxygen,
nitrogen, and neon abundances in the interstellar medium of 
\uks, and discuss it in the context
of the existence of primordial galaxies.


\section{Observations and data reduction}

\realfigure{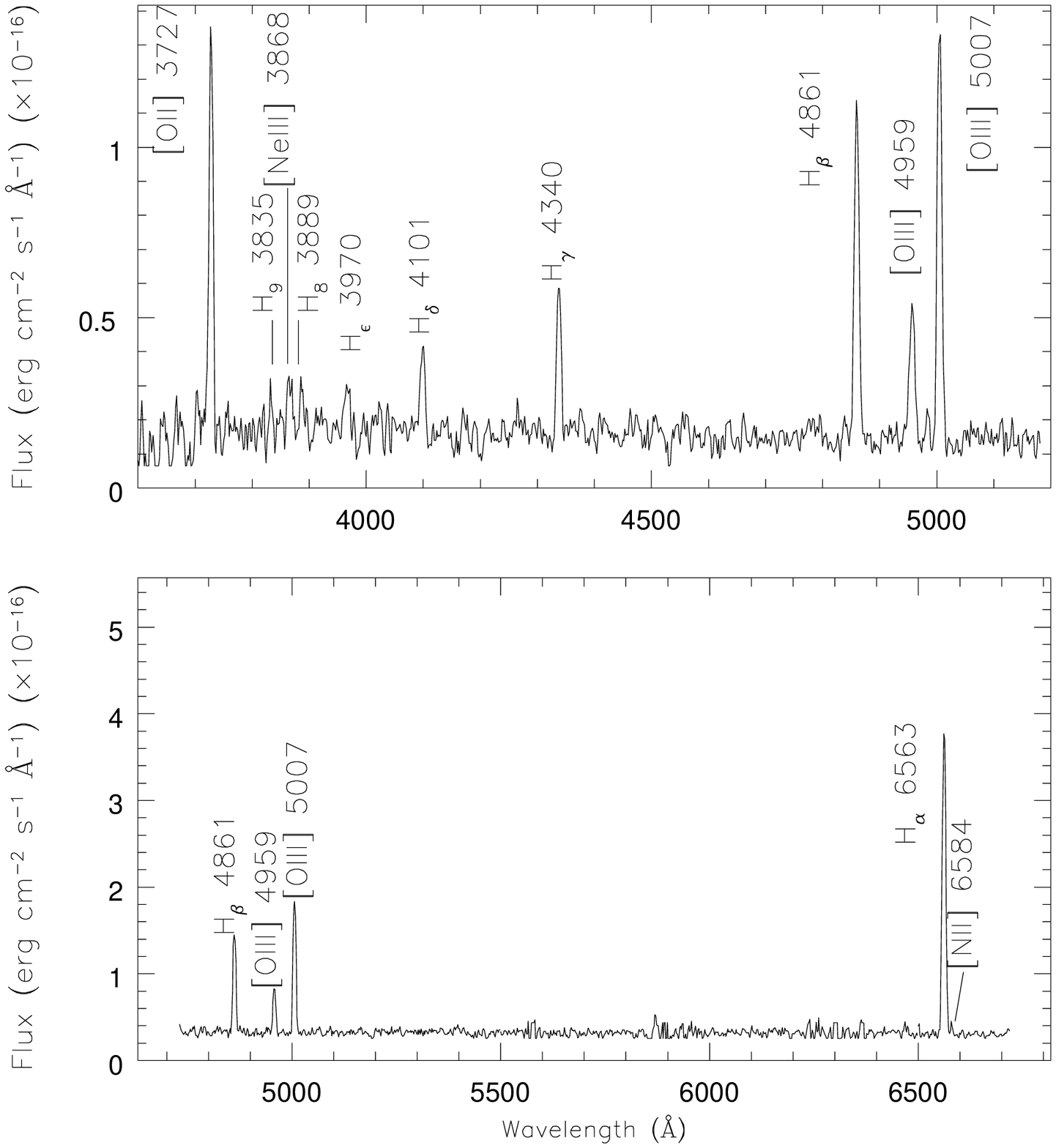}{
The combined spectra of the brightest \hii\ region of SagDIG (object
No.3 in Strobel et al. \cite{strobelEtal91}).  The upper panel shows
the line identification for the averaged flux-calibrated spectrum
obtained with grism \#7.  The lower panel shows the same except for
grism \#9.
}{f_idgr7+9}

Spectroscopic observations of SagDIG were carried out on 1999 August
10, at the ESO observatory in La Silla, Chile. The imaging
spectrograph \efosc\ attached to the ESO 3.6m telescope was
used to secure low-resolution spectroscopy of SagDIG. The camera
was equipped with a Loral/Lesser thinned, AR coated, UV flooded, \(
2048^{2} \) pixel CCD (ESO CCD \( \#40 \)).  The \( 15\mu \) pixels
cover \( 0\farcs 157 \) on the sky, with a resulting field of view of
\( 5\farcm 4\times 5\farcm 4 \).
Two grisms were employed to cover the wavelength ranges \( 3270\div
5240 \)~\AA\ (grism \#7) and \( 4700\div 6770 \)~\AA\ (grism \#9),
respectively. The CCD was used in \( 2\times 2 \) binning mode to
increase the \( S/N \) ratio, the resulting dispersion being \( \sim 2
\)~\AA/pixel for both grisms.

We observed the same \hii\ region studied by STM89, whose coordinates
are $\alpha=19$:$30$:$02.5$, $\delta=-17$:$41$:$26.1$ (J2000.0).
The night was stable and photometric, with a fairly constant seeing of
\( \sim 1\arcsec \) (FWHM) at H\( \alpha \). Spectroscopic data
comprise exposures of \( 6\times 1800 \) s with grism \#7 and \( 4\times
1800 \) s using grism \#9. In both cases, a slit width of \( 1\farcs 5 \)
was used. 
%
Spectrophotometric standards were observed during the night, selected
from the catalogs of Oke (\cite{oke90}) and Hamuy et al. (\cite{hamuy92},
\cite{hamuy94}).

Data reduction was performed using standard techniques in {\sc
iraf}\footnote{ IRAF is distributed by the National Optical Astronomy
Observatories, which are operated by the Association of Universities
for Research in Astronomy, Inc., under cooperative agreement with the
National Science Foundation.}.
The raw two-dimensional images were bias subtracted and trimmed.
Pixel-to-pixel variations were accounted for through division by 
a quartz lamp spectrum, after fitting a low-order polynomial along
the dispersion.  The splitting of the total exposure time in
\( 1800 \) s exposures allowed us to remove the
contamination from cosmic rays.
Wavelength calibration was obtained from He-Ar exposures taken
immediately before and after the object observations.  Absolute flux
calibration was achieved using standard stars, obtaining a final
accuracy better than \( 5\% \). 
One-dimensional spectra were extracted from the two-dimensional images
by summing up all the pixels enclosed in the large profiles of
H\({\alpha } \) and H\( {\beta } \); the background and sky lines
were subtracted by averaging a small region on either side of the
extraction window.

\tablecaption{Line flux ratios relative to $\mbox{H}{\beta}$ \label{t_ratios}}
\tablenotetext{
   \begin{list}{}{}
     \item[$^{\mathrm{a}}$] {possible blend with 3968 [\otre].}
     \item[$^{\mathrm{b}}$] {estimated maximum error (0.15) (see text).}
   \end{list}
}
\begin{deluxetable}{cllcc}
\tablehead{
\colhead{\(\lambda\)(\AA)} &
\colhead{Ident.} &
\colhead{Grism \#7} &
\colhead{Grism \#9} &
\colhead{Redd. corrected}
}
\startdata
3727&
 [\otwo] &
1.02\( \pm  \)0.20&
\nodata&
1.31\( \pm  \)0.21\\
3835&
H$_9$ &
0.08\( \pm  \)0.15\tablenotemark{b} &
\nodata&
0.10\( \pm  \)0.16\\
3868&
[\netre] &
0.13\( \pm  \)0.05&
\nodata&
0.16\( \pm  \)0.08\\
3889&
H$_8$ &
0.12\( \pm  \)0.15\tablenotemark{b}&
\nodata&
0.15\( \pm  \)0.16\\
3970&
H${\epsilon }$\tablenotemark{a}&
0.17\( \pm  \)0.15\tablenotemark{b}&
\nodata&
0.21\( \pm  \)0.16\\
4101&
H${\delta}$ &
0.19\( \pm  \)0.07&
\nodata&
0.23\( \pm  \)0.08\\
4340&
H${\gamma}$ &
0.36\( \pm  \)0.11&
\nodata&
 0.41\( \pm  \)0.11\\
4861&
H${\beta}$ &
1.00\( \pm  \)0.04&
1.00\( \pm  \)0.03&
1.00\( \pm  \)0.01\\
4959&
[\otre] &
0.48\( \pm  \)0.06&
0.46\( \pm  \)0.04&
0.46\( \pm  \)0.01\\
5007&
[\otre] &
1.35\( \pm  \)0.10&
1.34\( \pm  \)0.02&
1.30\( \pm  \)0.01\\
6563&
H${\alpha }$ &
\phantom{000}\nodata&
3.58\( \pm  \)0.03&
2.85\( \pm  \)0.05\\
6584&
[\ntwo] &
\phantom{000}\nodata&
0.15\( \pm  \)0.06&
0.12\( \pm  \)0.07 \\
\end{deluxetable}


Figure~\ref{f_idgr7+9} shows the combined spectra used for measurements.
The final resolution,
measured either on sky lines or comparison lamp lines, is \( 8.3 \)~\AA\
for grism \( \#7 \) and \( 9 \)~\AA\ for grism \( \#9 \).
Table \ref{t_ratios} presents the final results. From left to right,
the columns report the line wavelength and identification, the 
H\({\beta } \)-normalized fluxes for grisms \#7 and \#9, and the
reddening corrected fluxes (see below). 
Three emission lines ([\otre] $\lambda\lambda 4959, 5007$ and H${\beta}$)
are in common between the two grisms, showing no significant flux
offset.
In the worst case, that of the {[}\otre{]} \( \lambda 5007 \) line, 
the agreement is better than \( 5\% \).
A deblending procedure was applied to the case of the {[}\ntwo{]} \(
\lambda 6584 \) line to disentangle the contribution of the strong 
H\({\alpha } \).
%
The error budget was estimated from the r.m.s. of line ratios measured
on the different individual exposures.  A maximum error of 15\% on the
flux ratio was attached to our estimate when the line is not detected
in the single-exposure spectra.


The measured line ratios were corrected for the effects of 
reddening, measured by comparing the observed and the expected
hydrogen line ratios.
%
%
The theoretical H line ratios of Hummer \& Storey
(\cite{hummer}) were used, assuming an electron temperature 
\( T_{\rm e}\simeq \TeHummer \) K.
%
%
We set 
\(R=I_{\lambda }/I_{\rm H\beta } \), 
\( R_{0}=I_{\lambda, 0}/I_{\rm H\beta, 0} \) and 
\( \varphi (\lambda )=[f(\lambda )-f(H\beta )] \) so that 
\( C=[\log R_{0}/R]/\varphi (\lambda ) \).  
The reddening constant was then computed using the 
Cardelli et al. (\cite{cardelli}) extinction law,
%
where $f(\lambda)=<A(\lambda)/A(V)>$, and we assumed $R_V=3.1$.
 We obtained 
\( C=\cred\pm \ecred \) from the H\( {\alpha } \)/H\( {\beta } \) ratio, and
\( C = \credGamma \pm \ecredGamma \) from the 
H\( {\beta } \)/H\( {\gamma } \) ratio. 
As an alternate route, one can plot \( \log (R_{0}/R) \) vs. \(
\varphi (\lambda ) \) and then find \( C \) with a linear regression
on the data. A weighted least-square fit on the first three H lines
yields 
\( C= \credRegression  \pm \ecredRegression \).  
Thus the value of the reddening constant is
essentially fixed by the more accurate 
H\( {\alpha } \)/H\({\beta } \) ratio, 
and \( C=\cred\pm \ecred \) will be adopted in the rest of this paper.
The reddening corrected flux ratios are listed in
Table~\ref{t_ratios}. When the same line is present in both the red
and blue spectra, a weighted mean is computed before the correction is
applied.  In order to compute the final errors, the error contribution
from the reddening correction
was added in quadrature to the error on the flux.

\section{Results}

Chemical abundances were derived from these measurements using the {\sc
iraf} \texttt{nebular} package (Shaw \& Dufour \cite{sd95}).
The abundances were calculated assuming a constant electron temperature
and density throughout the nebula.
For optical spectra covering the observed wavelength range,
the electron temperature should be obtained through $T_{\rm e}$-sensitive
line ratios, e.g. from
the ratio of {[}\otre{]} lines \( R_{4363}=[I(4959)+I(5007)]/I(4363) \)
(see Eq.~2.3 in Seaton \cite{seaton75}). However, the \( \lambda 4363
\) line was not detected even in our co-added spectra, so we 
had to use the so-called empirical method, which gives the total
(O/H) abundance in terms of the \( \remp =
[I(4959)+I(5007)+I(3727)]/I(H\beta ) \) parameter (Pagel et
al. \cite{pagel_etal79}; Edmunds \& Pagel
\cite{edm_pagel84}). 

\realfigure{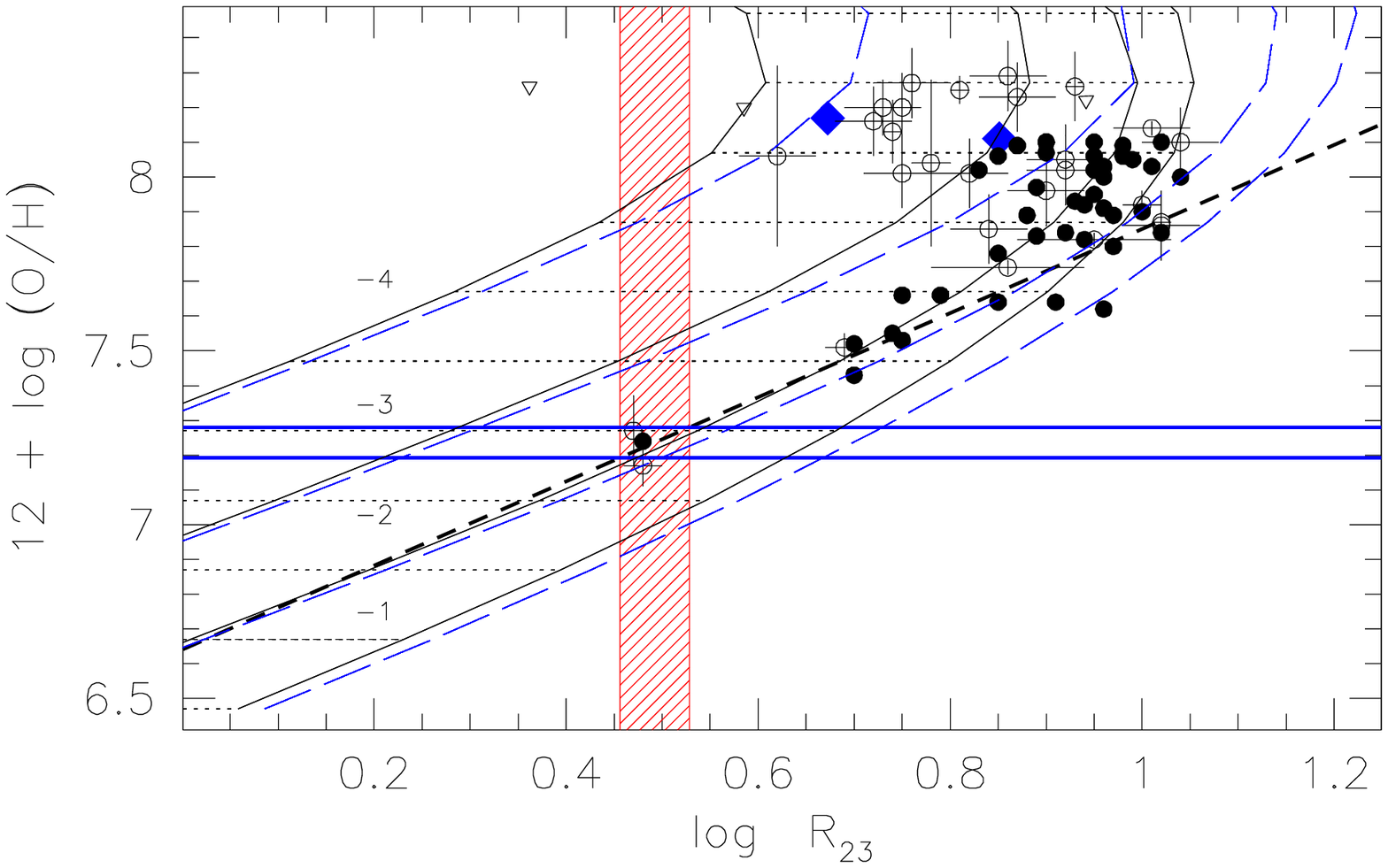}{
(O/H) determination for SagDIG using the empirical method based on
$\remp$. The $1\sigma$ range of variation of our value of $ \log(\remp)$
is shown by the vertical band. The empirical relation of Skillman
(\cite{skillman89}) is represented by the inclined heavy short-dashed
line.  The horizontal solid lines show the range of uncertainty of $\log
\rm (O/H)$.
For comparison, 
values of \( \log \rm (O/H) \) are plotted against the value
of \( \log \remp \) for the sample of \hii\ regions from 
D{\'{\i}}az \& 
P{\' e}rez-Montero (\cite{dpm00}) ({\it open circles}), Oey
\& Shields (\cite{oey00}) ({\it filled diamonds}),  Kennicutt et
al. (\cite{kennicuttEtal00}) ({\it open triangles}), and McGaugh
(\cite{mcga91}) ({\it filled circles}). Only objects more metal poor
than $12+\log(\rm O/H)=8.3$ are plotted.
A number of {\sc cloudy} simulations are also shown.
The thin solid curves represent the 
{\changed theoretical} relations obtained by
varying the ionization parameter from $\log U = -1 $ to $\log U = -4$
(see labels), and by varying [m/H] from $-0.4$ to $-2.4$ 
(top to bottom, {\it dotted lines}). An input ionizing spectrum from a 
\( T_{\rm eff}=40\,000\, \rm K \) star was assumed. The effect of
changing the input spectrum is shown by the long-dashed curves, 
which were calculated for \( T_{\rm eff}=50\,000\, \rm K \).
}{f_emp_meth}

Oxygen abundances are plotted against $\remp$ in
Figure~\ref{f_emp_meth}, using literature samples of \hii\ regions
spanning a wide range of abundances (Oey \& Shields \cite{oey00}; 
D{\'{\i}}az \& P{\' e}rez-Montero \cite{dpm00};
Kennicutt et al. \cite{kennicuttEtal00};
McGaugh \cite{mcga91}). 
In the same diagram, we show a set of {\sc cloudy} (Ferland et
al. \cite{ferlandEtal98}) simulations, obtained assuming that the
ionizing source is a single star with a Mihalas (\cite{mihalas72})
continuum, for \( 4 \) choices of the ionization parameter (\( \log
U=-1,\, -2,\, -3 \) and \( -4 \)) and varying the metallicity from
$12+\log \rm (O/H)=6.47$ to $12 + \log\rm (O/H)=8.87$ (solar composition).
The effects of a change in the temperature of the
ionizing star were checked by running the simulations for both  
\( T_{\rm eff}=40\,000\, \rm K \) and 
\( T_{\rm eff}=50\,000\, \rm K \). 

This figure, an updated version of Fig.~2 of Skillman (\cite{skillman89};
hereafter S89) and Fig.~12 of McGaugh (\cite{mcga91}),
shows that, for abundances $12+\log(\rm O/H)\leq 8$,  
$\remp$ depends both on the metallicity and the ionization
parameter of the nebula.
%
The S89 relation corresponds to
the lower branch of the \( \log U=-2 \) models
%
and to the most metal-poor objects in the data samples,
irrespective of the effective temperature of the ionizing source.

Several authors have attempted to include both the effects of 
metallicity and ionization parameter 
in their empirical calibration methods. 
According to McGaugh (\cite{mcga91}), 
the parameter \( \log (\rm [\otre]4959,5007/[\otwo]3727) \)
is useful to constrain $U$. 
%
%
%
More recently, Pilyugin (\cite{pilyugin00}, \cite{pilyugin01a},
\cite{pilyugin01b}) has provided useful formulae to implement McGaugh
corrections into the empirical method. 
Pilyugin (\cite{pilyugin01b}; hereafter P01) transformed the
$\log(\rm O/H)=f(\remp,\log U)$ relation into a $\log(\rm
O/H)=h(\remp,R_3)$ relation, where 
$R_3=\log (\rm [\otre]4959,5007/H\beta$). In this case he gives the
equation of the surface, which is:
\[
12 + \log (\rm O/H) =  6.35 + 3.19 \times \log(\remp) - 1.74 \times \log(R_3)
\]
These revised methods may have problems at low values of 
the abundance and the ionization parameter.
For example, the dwarf galaxies ESO\,380-27 and GR\,8 have similar $\log
(\remp)$, but $\log (R_3)=0.67$ for the former, and $\log
(R_3)=0.43$ for the latter. The low excitation parameter then implies
that GR\,8 should have a lower $\log U$, implying a higher oxygen
abundance. Indeed, we find $12+\log(\rm O/H)=7.58$ for the first, and
$12+\log(\rm O/H)=7.83$ for the second galaxy. The corrected empirical
method would then predict that GR\,8 should have an oxygen abundance
$0.25~\rm dex$ higher than ESO\,380-27, but in reality the 
measured abundance from
the temperature determinations is $0.10~\rm dex$ {\em lower} in GR\,8
than in ESO\,380-27 (see Kunth \& Sargent \cite{kunthSargent83}; 
Skillman et al. \cite{skillmanEtal88}).
%
%
%
\realfigure{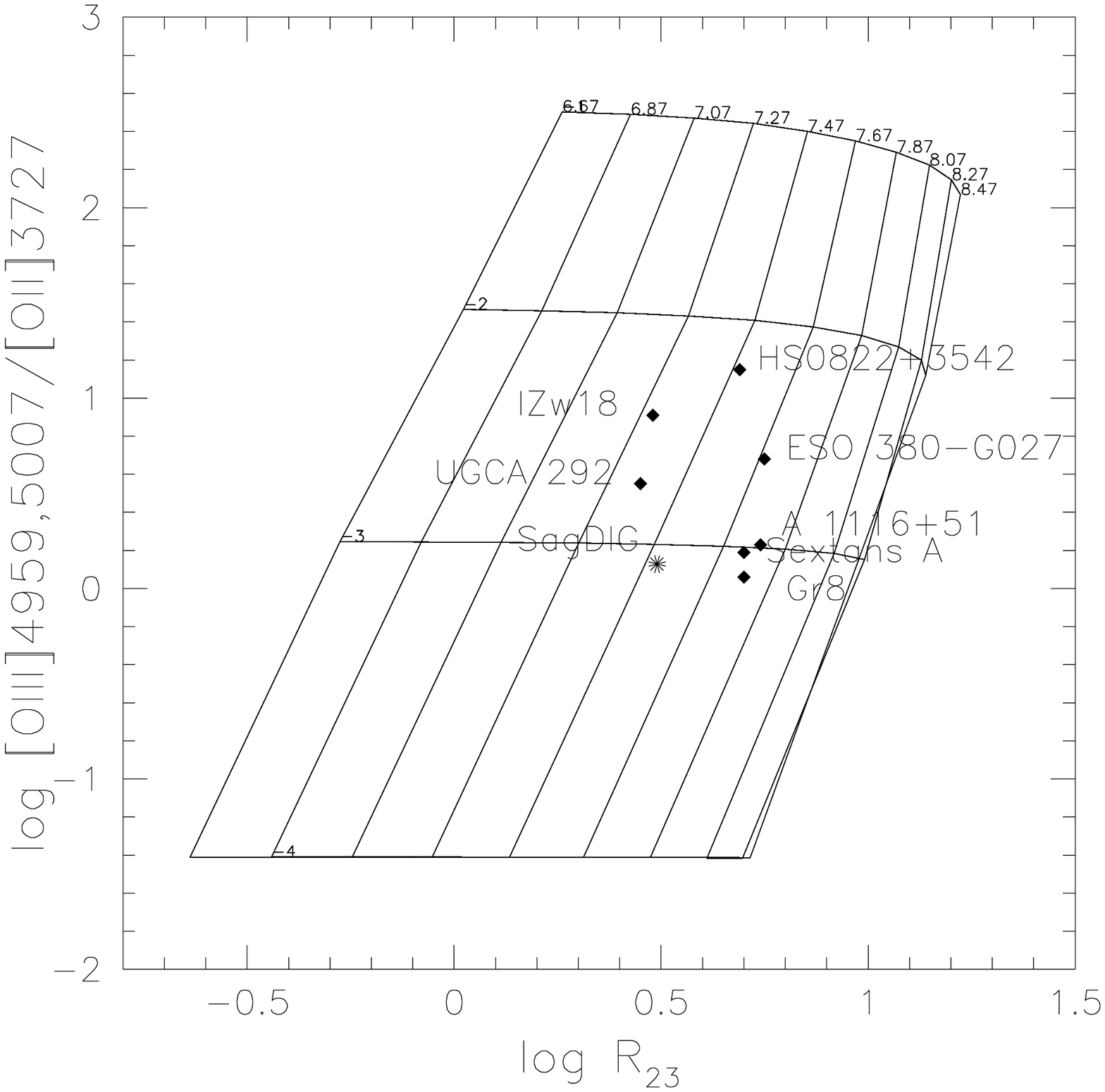}{
A comparison of the observed values of \( \remp\ \) and the
[\otre]/[\otwo] ratio for the most metal-poor star-forming dwarfs with
a model grid obtained from {\sc cloudy} simulations, for $T_{\rm
eff}=50000~\rm K$.  The data are from Skillman (\cite{skillman89}), 
McGaugh (\cite{mcga91}), Van Zee (\cite{vzee+99}), and Kniazev et al. (\cite{kniazev00}).}{f_mcgaugh91}
%
Given all these uncertainties,  
we used both the original implementation of the empirical method
(as in STM89) and more recent methods to provide a range of
[O/H] for SagDIG. 
High $S/N$ spectra taken at 10-m class telescopes will be needed to 
place better constraints on the SagDIG abundance, by allowing the
measurement of the $\lambda 4363$ [\otre] line.

Therefore, we firstly used the relative O fluxes 
listed in Table~\ref{t_ratios} to obtain
\( \remp = \rduetre \pm \erduetre \) 
(i.e. \( \rduetremin \leq \log \remp \leq \rduetremax \)). 
The calibration of S89 
%
gives \( \dodOHrMin \leq 12 + \log (\rm O/H)\leq \dodOHrMax \), where
the range in variation has been obtained by propagating the error 
on \( \remp \).
An application of the empirical calibration therefore indicates a
\emph{very low oxygen abundance} for the Sagittarius dwarf irregular
galaxy.  This value can be directly compared to the abundance
obtained by STM89 using the same calibration.  The new abundance
estimate is lower by $\sim0.2$ dex, and the difference is real and
traced to the new flux measurements.  Since our measurements from
the red and blue \efosc\ spectra are perfectly consistent, whereas the
measurements of STM89 presented a \( \sim 40\% \) internal scatter, we
are confident in our revised fluxes (indeed, using only the STM89 {\it
red} spectrum for the
\( \lambda \lambda 4959,5007 \) fluxes, we would obtain a value
\( 12 + \log (\rm {O/H})=7.22 \), in perfect agreement with ours). 

Secondly, the revisions of the empirical method were taken into
account by plotting in Fig.~\ref{f_mcgaugh91} the [\otre]/[\otwo]
ratio against the \( \remp\ \) for a subsample of very metal-poor
dwarfs (see Fig.~10 in McGaugh \cite{mcga91}).  The data are
overplotted onto a model grid derived from the same {\sc cloudy}
models as in Fig.~\ref{f_emp_meth}.
In the case of SagDIG, we have $\log (\rm
[\otre]4959,5007/[\otwo]3727)\simeq0.13$, which implies $\log U \approx -3$. 
%
The position of SagDIG with respect to the lines of constant O
abundance indicates a metallicity intermediate between those of
I\,Zw\,18 and GR\,8, and close to that of HS0822+3542, for which
Kniazev et al. (\cite{kniazev00}) give $12+\log (\rm O/H) = 7.35$.  Thus, taking into account
{\it in a differential way} the information provided by the ionization
indicator [\otre]/[\otwo], we find $12+\log (\rm O/H) \simeq \dodm91$.

Finally, we evaluated the O abundance by using {\sc cloudy} to
optimize the fit to the observed line ratios. The best-fit is obtained
for a value $12+\log (\rm O/H) = 7.41$, i.e. 1/29 solar, with an
electron temperature $T_e \sim 16\,000$ K and $\log U \simeq -2.8$.
Not surprisingly, this result is consistent with that inferred from
Fig.~\ref{f_mcgaugh91}.  


%

To proceed further to estimate the N and Ne abundances in SagDIG, we
need making some considerations (and assumptions) on the electron
temperature and density in the nebula. 
A loose upper limit on the electron temperature is set by non-detection
of the \( \lambda 4363 \) {[}\otre{]} line.  Experiments simulating faint
emission lines yield a $2\sigma$ upper limit of \( 0.09 \) on the flux
of the \(\lambda 4363 \) line relative to H$\beta$, which implies
$T_{\rm e} < 34\,000$ K according to the Seaton's (\cite{seaton75})
relation.
%
%
%
A temperature $T_{\rm e}=20\,000 ~ \rm K$ is needed in the
\texttt{nebular} calculations to obtain an (O/H) value compatible with
the results from the empirical method (assuming \( N_{\rm e}=1000\ \rm
cm^{-3} \)).
In fact, using these values we obtain
\(N(\rm {O^{+}})/N(\rm H^{+}) = ( \oiiHii  \pm \eoiiHii ) \times 10^{-5} \) 
and
\(N(\rm {O^{++}})/N(\rm H^{+}) = ( \oiiiHii  \pm \eoiiiHii) \times 10^{-5} \), 
yielding  
\( (\rm {O^{+}}+\rm {O^{++}})/\rm {H^{+}} = 
( \oiioiii \pm \eoiioiii) \times 10^{-5} \), 
(i.e. $12 + \log \rm [(O^+ + O^{++})/H^+] = 7.26$).  
%
%
%
This choice of physical parameters predicts
\( R_{4363} \simeq 37 \), implying that the \( \lambda 4363 \) line
flux ratio should be \( \simeq 0.05 \), definitely lower than the
minimum normalized flux that we have been able to measure. 
%

Yet, the low surface brightness of the nebula seems to suggest 
a lower electron temperature and ionization parameter, in accord with 
the above {\sc cloudy} model calculations. 
%
In view of these uncertainties, we adopt 
$T_e = 18\,000 \pm 2\,000$ for electron temperature, implying
$12+\log(\rm O/H) = 7.37^{+0.13}_{-0.11}$ for the O abundance in SagDIG, 
in accord with the different empirical approaches discussed above. 
For the nitrogen and neon ionic abundances in \uks,
\( (\rm {N^{+}/H^{+}})  = ( 0.062 \pm 0.039)\times 10^{-5} \)
and
\( (\rm {Ne^{++}/H^{+}})= ( 0.284 \pm 0.134 )\times 10^{-5} \) are derived.
The total abundances can be computed following Peimbert \& Costero
(\cite{peimbertCostero69}) (see also van Zee \cite{vzee98}), 
obtaining $12+\log(\rm N/H)= 6.00^{+0.28}_{-0.98}$
 and $12+\log(\rm Ne/H) = 6.86^{+0.20}_{-0.38}$.  This
also implies $\log(\rm N/O)=-1.36^{+0.30}_{-1.76}$ and $\log (\rm
Ne/O)=-0.50^{+0.22}_{-0.51}$.  
The quoted uncertainties for the N and Ne abundances reflect the
measurements errors and are larger than those related to the temperature range.
Notwithstanding the above uncertainties, this represents the first
measurement of N and Ne abundances in \uks.

\section{Discussion}


Our results confirm that SagDIG is indeed one of the most metal poor
galaxies, having abundances just $50\%$ higher than I~Zw~18.
%
%
%
In the Local Group, only 
Leo~A has a comparably low metallicity.
%
The nitrogen abundance of SagDIG is consistent with that of
I\,Zw\,18 (Izotov \& Thuan \cite{izotov98}).
For Leo~A, van Zee et al. (\cite{vzee+99}) report a nitrogen abundance
\( \log {\rm N/O} = -1.5 \pm 0.1 \), similar to other low-metallicity 
galaxies. 
%

Answering the question about whether SagDIG and Leo\,A are young
galaxies rests on the results of resolved photometry.  
For SagDIG, ground based observations indicate the presence of an
intermediate-age stellar population in addition to the young stars
(Karachentsev et al. \cite{kara_etal99}; Momany et
al. \cite{yaz_sagdig}), but the presence of a really old component is
still unproven.
As is the case of Leo\,I, where wide-field observations have revealed
an old HB (Held et al. \cite{held_etal00}, \cite{held_etal01}), deep
{\sc hst} observations of SagDIG may eventually answer the original
question of how old this metal-poor galaxy is.

In Leo\,A, Tolstoy et al. (\cite{tolstoy_etal}) could not rule out the
presence of such a population in their {\sc hst} photometry, although
it must be a minor component.  Recently, Dolphin et
al. (\cite{dolp+02}) have detected a few candidate RR Lyrae variable
stars, indicating the presence of an old (globular cluster like)
stellar population accounting for at least 0.1\% of the galaxy's
luminosity.

{\changed Finally, it is important to check that 
the low metal content of SagDIG is the result of normal chemical
evolution, and not, for example, of gas stripping by a massive
neighbor. Indeed, its position at the margins of the Local Group seems
to ensure self-regulated evolution, an idea that we reinforce by
the following simple considerations of chemical evolution.  }


In the closed-box model, the expected ISM oxygen
abundance is given by 
\( (\rm O/H) = (O/H)_{y} \ln ({1}/\mu) \)
(Searle \& Sargent \cite{ssargent72}).
A {}``standard{}'' value for the yield of oxygen, \( \rm (O/H)_{y} \),
is \( 2/3 \) of the solar abundance, i.e. \( \rm (O/H)_{y}=4.9\times
10^{-4} \) (e.g. Pagel \cite{pagel97}); and \(\mu =G/(G+S) \) is the
ratio of the gas mass \( G \) over total mass (\( S \) is the mass in
stars).

The gas mass given by Young \& Lo (\cite{youngLo97}) is \( 1.3\times
10^{7}\, \rm \textrm{M}_{\odot } \), and the mass in stars and
remnants, computed by integrating the SFR given by Karachentsev et
al. (\cite{kara_etal99}) in three age intervals, is \(
2\times 10^{6}\, M_{\odot } \).
We thus obtain \( \mu =0.86 \), a value that in the closed-box model
implies \( 12+\log (\rm O/H)=7.87 \), higher than the measured one.
We would need \( \mu =0.97 \) to reproduce our observation.  The
observed lower O abundance could be explained if part of the enriched
gas is lost into the intergalactic medium.
With some algebra, we find that the mass lost should be
\( \Delta G=0.11\times G=1.5\times 10^{6}\, \rm \textrm{M}_{\odot }
\).  \
According to Karachentsev et al. (\cite{kara_etal99}), the SFR
during the last \( 200\, \rm Myr \) was \( \sim 10 \) times higher than
in the past, with a further enhancement during the last \( 50\, \rm Myr
\). If we assume that the mass loss occurred during the last \( 200 \)
million years, then we obtain a mass outflow of \( \dot{M}=0.01\, \rm
\textrm{M}_{\odot }\, yr^{-1} \), while \( \dot{M}=0.03\, \rm
\textrm{M}_{\odot }\, yr^{-1}
\) if the mass outflow occurred during the last \( 50 \) million
years. These are not unreasonable values, since for example outflows
of the order of \( 0.2\div 2\, \rm M_{\odot }\, yr^{-1} \) are
observed in the case of NGC~1705 (Meurer et al. \cite{meurer_etal98}),
{\changed another isolated dwarf galaxy.}


Using the higher value of the gas fraction, 
we find an {average} metallicity for
the \emph{stellar} component which is in agreement with the values
derived from broadband photometry.  Using Eq.~(8.7) of Pagel (1997),
an average oxygen abundance of \( 0.01 \) solar is obtained for
the stars. For solar-scaled metal abundances, this would mean
a metallicity \( \rm [Fe/H]=-2 \), 
%
%
which agrees with the stellar metallicity measured by 
Momany et al. (\cite{yaz_sagdig}), [Fe/H]$\simeq -2.1$. 

Our conclusion that a very metal-poor ISM is present in SagDIG goes in
the direction of the accumulating evidence that very low oxygen
abundances can be maintained throughout the life of a
dwarf galaxy, even if it is not experiencing its first star formation
episode,
{\changed and without invoking special environmental conditions.}
It is fairly common that galaxies once believed to contain
only a young population actually reveal an old halo when modern data
are obtained (see Kunth \& {\"O}stlin
\cite{kunthOstlin00} for a review; also Saviane et
al. \cite{saviane_etal01}).


\acknowledgements 
We thank Michael Sterzik for his precious help with the observations,
and Stefano Ciroi for sharing with us his expertise in {\sc cloudy}.
We also thank the referee, Dr. C. Leitherer, for helpful comments. 



\end{document}